\documentclass[aps,prx,superscriptaddress,twocolumn,balancelastpage,amsmath,amssymb,longbibliography=false]{revtex4-2}

\usepackage{textcomp}  % in the preamble

\usepackage{graphicx}
\usepackage[dvipsnames]{xcolor}
\usepackage[colorlinks,linkcolor=black,citecolor=black,urlcolor=black,filecolor=black]{hyperref}
\usepackage{amsmath, amssymb, amsthm}
\usepackage{wasysym}
\usepackage{lipsum}
\usepackage{outlines}
\usepackage{enumerate}
\usepackage{setspace}
\usepackage{tabularx}
\usepackage{xcolor}
\usepackage[export]{adjustbox}
\usepackage{tikz}
\usepackage{verbatim}
\usepackage{ulem}
\usetikzlibrary{decorations.pathreplacing}
\usetikzlibrary{arrows,arrows.meta,calc,shapes.geometric,shapes.misc}
\tikzset{
	>=stealth',
	help lines/.style={dashed, thick},
	important line/.style={thick},
	connection/.style={thick, dotted},
}
\DeclareMathAlphabet{\mymathbb}{U}{BOONDOX-ds}{m}{n}
\usepackage{dsfont}
\usepackage{braket}
\hypersetup{
    pdfstartview={FitH}, 
    colorlinks=true, 
    linkcolor=blue, 
    citecolor=blue, 
    filecolor=magenta,
    urlcolor=blue
}

\begin{document}

\title{
% An Efficient Method for Constructing a\\Neutral Atom Cluster State using an Optical Cavity
Efficient construction of fault-tolerant neutral-atom cluster states
}

\author{Luke M. Stewart}
\affiliation{Department of Physics, MIT-Harvard Center for Ultracold Atoms and Research Laboratory of Electronics,
Massachusetts Institute of Technology, Cambridge, MA 02139, USA}
\affiliation{Department of Physics, Harvard University, Cambridge, MA 02138, USA}

\author{Gefen Baranes}
\affiliation{Department of Physics, MIT-Harvard Center for Ultracold Atoms and Research Laboratory of Electronics,
Massachusetts Institute of Technology, Cambridge, MA 02139, USA}
\affiliation{Department of Physics, Harvard University, Cambridge, MA 02138, USA}

\author{Joshua Ramette}
\affiliation{Department of Physics, MIT-Harvard Center for Ultracold Atoms and Research Laboratory of Electronics,
Massachusetts Institute of Technology, Cambridge, MA 02139, USA}

\author{Josiah Sinclair}
\affiliation{Department of Physics, MIT-Harvard Center for Ultracold Atoms and Research Laboratory of Electronics,
Massachusetts Institute of Technology, Cambridge, MA 02139, USA}

\author{Vladan Vuleti\'{c}}
% \email{Contact author: vuletic@mit.edu}
\thanks{Contact author: \href{mailto:vuletic@mit.edu}{vuletic@mit.edu}}
\affiliation{Department of Physics, MIT-Harvard Center for Ultracold Atoms and Research Laboratory of Electronics,
Massachusetts Institute of Technology, Cambridge, MA 02139, USA}

\date{\today}

\begin{abstract}

%Deterministic two-qubit gate errors remain a major obstacle to scalable fault tolerance in neutral atoms.
%
Cluster states are a useful resource in quantum computation, and can be generated by applying entangling gates between next-neighbor qubits. Heralded entangling gates offer the advantage of high post-selected fidelity, and can be used to create cluster states at the expense of large space-time overheads. We propose a low-overhead protocol to generate and merge high-fidelity many-atom entangled states into a 3D cluster state that supports fault-tolerant universal logical operations. Our simulations indicate that a state-of-the-art high-finesse optical cavity is sufficient for constructing a scalable fault-tolerant cluster state with loss and Pauli errors remaining an order of magnitude below their respective thresholds. This protocol reduces the space-time resource requirements for cluster state construction, highlighting the measurement-based method as an alternative approach to achieving large-scale error-corrected quantum processing with neutral atoms.

\end{abstract}

\maketitle

\section{Introduction}
Trapped neutral atoms are a promising platform for quantum computing. They combine scalability with long coherence times ($> 1$ s), and high fidelity single-qubit operations ($99.9$\%) and measurement ($99.99$\%) \cite{Manetsch2024_6100,Bluvstein2022_transport,Jenkins2022_coherence,Ma2023_erasure}.
Nevertheless, quantum error correction (QEC) is necessary to achieve sufficiently low error rates and deep circuits required for applications of interest \cite{Beverland2022_resource,Zhou2024_algoFT}. Recent experimental advancements have enabled the first QEC protocols with neutral atoms. 
In particular, neutral-atom qubits can be dynamically reconfigured within a zoned architecture while preserving coherence, \cite{Bluvstein2022_transport,Bluvstein2024_logical,Manetsch2024_6100} mid-circuit measurements can extract syndrome information without affecting data qubits, and dominant loss and leakage errors can be converted into known erasure errors and utilized in decoding \cite{Singh2023_midcircuit,Norcia2023_midcircuit,Lis2023_midcircuit,Ma2023_erasure,Reichardt2024_logical,Chow2024_erasure,Muniz2025_repeatedreuse,Zhang2025_logicalerasure,Baranes2025_loss}.
Together with early demonstrations of fault-tolerant state preparation, error suppression scaling with surface code distance, classically non-trivial circuits executed with logical qubits, and mechanisms for coherence-preserving qubit replenishment, these results lay the foundation for large-scale QEC architectures based on neutral atoms \cite{Bluvstein2024_logical,Reichardt2024_logical,Li2025_reloading,Chiu2025_Reloading,Bluvstein2025_architectural}.

The threshold theorem \cite{Knill1998_threshold} states that the logical error rate of an error-correcting code can be made arbitrarily small by scaling up the system as long as the total physical error rate is below some threshold.
Thresholds can range from $10^{-6}$ \cite{Luo2020_threshold} to $10^{-2}$ \cite{Fowler2012_surface,Stephens2014_surface}.
Reducing physical error rates below $10^{-2}$ is a major focus in neutral-atom experiments, with significant progress made to date.  Currently, the limiting factor is the two-qubit gate fidelity, which has recently surpassed $99.5\%$ in certain species \cite{Evered2023_gate,Finkelstein2024_gate,Muniz2024_gate}.
However, as long as total physical error rates are at or near threshold, practical implementations of error correction remain a challenge due to the large number of physical qubits required. 

Measurement-based quantum computation (MBQC) is an alternate paradigm of computation originally developed for photonic platforms, which lack deterministic interactions.
Instead, heralded probabilistic gates are used to prepare a large entangled state called a ``cluster state", and the computation takes place by successive rounds of single-qubit gates and measurements \cite{Raussendorf2001_owqc,Raussendorf2003_mbqc,Briegel2009_mbqc,Schleier2024_graph}. The well-established topological quantum error correction scheme from Raussendorf \textit{et al.} \cite{Raussendorf2007_mbqc} achieves universal fault-tolerance using a three-dimensional cluster state geometry known as a Raussendorf-Harrington-Goyal (RHG) cluster state.
Numerical simulations of this approach have shown that the threshold for each error source including preparation, one- and two-qubit gates, and measurement  is as high as $0.67\%$ \cite{Raussendorf2007_mbqcnewj}, similar to the best circuit model thresholds \cite{Fowler2012_surface,Stephens2014_surface}. Unlike the circuit-based model, the MBQC approach offers an inherent resilience against qubit losses because the quantum information is teleported to new qubits at each syndrome extraction cycle \cite{Baranes2025_loss}. The approach accommodates heralded, probabilistic entangling gates which can achieve a post-selected gate fidelity limited only by measurement errors \cite{Borregaard2015_entanglement}. The threshold for entanglement failure between qubits is 14.5\% \cite{Auger2018_mbqc}.  We consider a method to efficiently create neutral-atom cluster states well below both of these thresholds.
% proving that universal fault-tolerant QC is possible without a high-fidelity deterministic two-qubit gate.
% The MBQC approach differs from the circuit model because 

% The errors that propagate to the cluster state largely stem from incorrect measurement outcomes during heralding, which in neutral atoms is several orders of magnitude higher fidelity than current Rydberg gates. (CITE)

% For the same number of physical qubits, the code distance of the MBQC cluster state will be less than a gate-based surface code implementation.  In terms of logical failure rates, however, the asymmetric error between the two approaches compensates for the lesser distance, meaning MBQC remains a viable alternative to the gate-based method at modern Rydberg gate fidelities. (Appendix details?)

The challenge in MBQC is preparing large cluster states efficiently using high-fidelity heralded entangling gates.
A `divide-and-conquer' method developed in \cite{Duan2005_divide} involves first generating finite-sized entangled states (``resource states") as a preliminary step. The cluster state is constructed by merging the resource states together with heralded two-qubit C-phase gate attempts, repeated until a successful entanglement ``edge" is formed.
The resource states provide overhead to accommodate failures of the heralded C-phase gate in this merge process; a gate with low success probability will require large resource states to prevent failed edges in the cluster state.
The imperfect success probability of the heralded gate leads to time and/or spatial overheads that are polynomial in the size of the resource state \cite{Duan2005_divide}.

\begin{figure}[h]
\includegraphics[width=\linewidth]{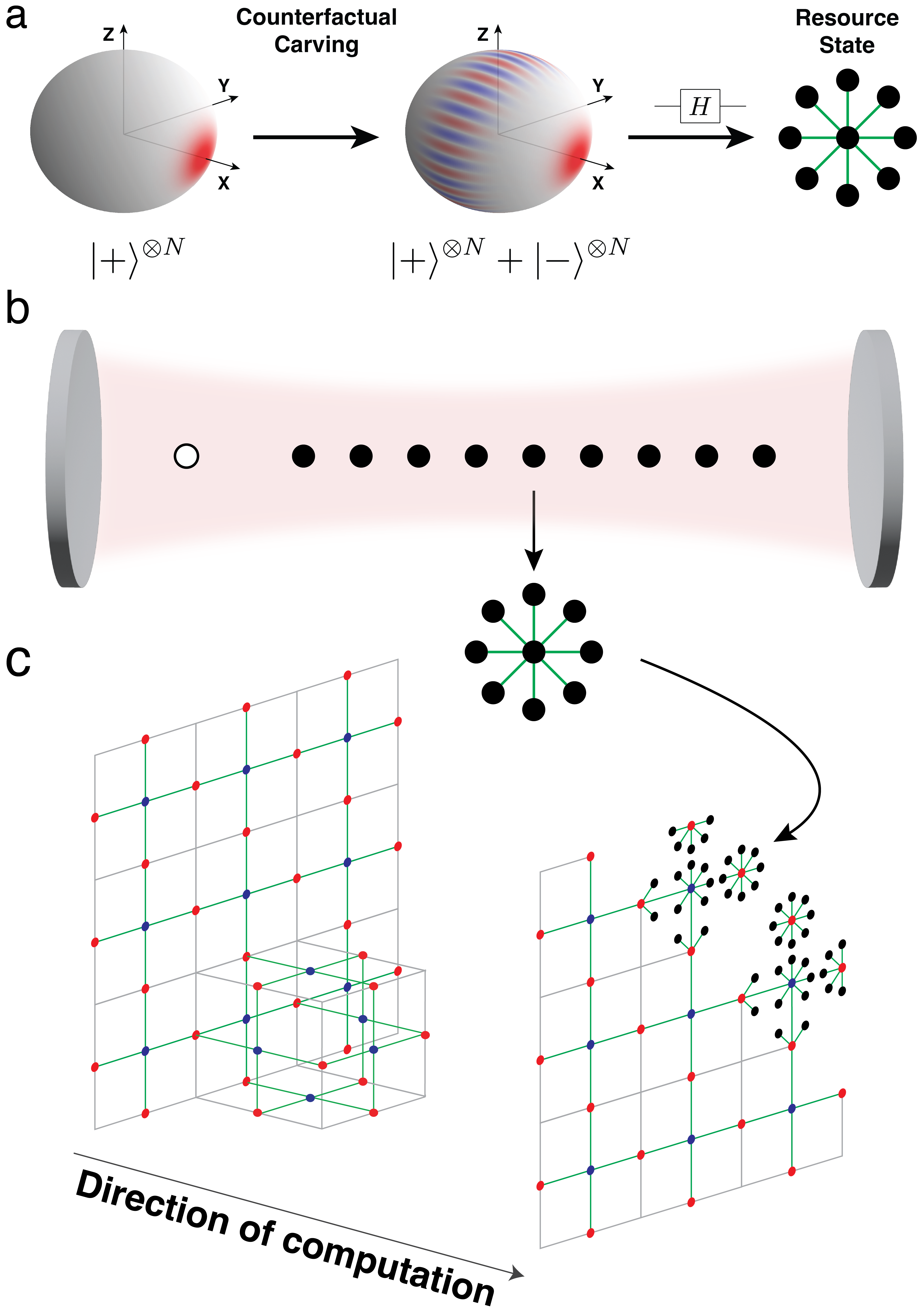}
\caption{\textbf{Cavity-based cluster state construction.} 
\textbf{a.} Counterfactual carving generates multipartite entangled resource states by ``carving" away unwanted Dicke levels from the initial coherent spin state. \textbf{b.} A high-finesse optical cavity is used to generate large entangled states of atoms (resource states) in a heralded procedure. \textbf{c.}  These states are then merged into a 3D topologically-protected fault-tolerant cluster state by a heralded, cavity-mediated gate.
}
\label{fig:overview}
\end{figure}

\begin{figure}[h]
\includegraphics[width=\linewidth]{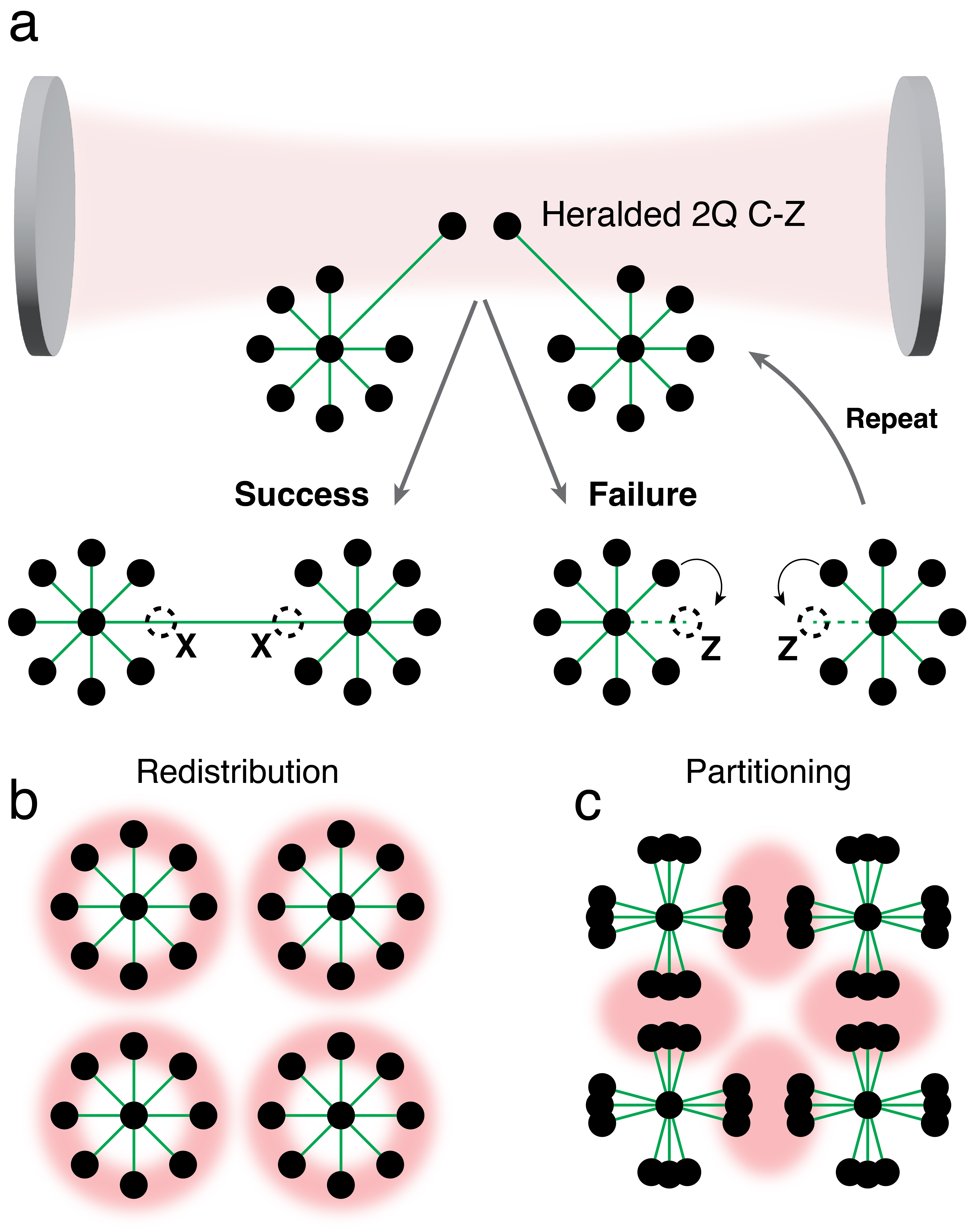}
\caption{\textbf{Protocols for constructing fault-tolerant cluster states using star-graph resource states. a.} The same cavity that supports counterfactual carving can also be used to entangle leaf qubits from neighboring resource states in a heralded manner.  Following a heralded success, the two leaf qubits are measured in the $X$ basis.  Following a heralded failure, the two leaf qubits are removed via $Z$ basis measurements and the operation is attempted again. \textbf{b, c.} Two strategies for merging resource states, redistribution and partitioning, differ based on whether a given leaf qubit is free for use on any bond (\textbf{b.}) or assigned to be used towards creating a particular bond of the cluster state (\textbf{c.}).}
\label{fig:construction}
\end{figure}

In the present work, we propose a neutral-atom measurement-based quantum computation scheme to achieve fault-tolerance with reduced time and space overheads. 
At the heart of our proposed method is a novel protocol for generating resource states in a single step, simplifying the construction process.  A high-finesse optical cavity is used to entangle an ensemble of neutral atoms in a heralded manner \cite{Ramette2024_counterfactual}, creating a resource state (``counterfactual carving", Fig. \ref{fig:overview}a).
To construct the cluster state from resource states, we propose to use a heralded, cavity-mediated C-phase gate \cite{Borregaard2015_entanglement} with small post-selected infidelity (Fig. \ref{fig:construction}a).  Our simulations indicate that, under realistic assumptions about cavity cooperativity and resource state size, we can construct a fault-tolerant cluster state that is at least one order of magnitude below error correction thresholds.  
Our scheme leverages the unique strengths of the neutral-atom platform, particularly its scalability, capacity for high-fidelity measurement and support for efficient heralded many-body entangling gates---all required for MBQC---while reducing the typical overhead in resource state generation.

The paper is organized as follows.  We first present our cavity-based method of preparing star-type many-body-entangled resource states in a single step based on counterfactual carving. We then show how to construct a cluster state from those resource states. Finally, we discuss the experimental feasibility of neutral-atom MBQC and show that our proposal for a fault-tolerant neutral-atom quantum computer can be implemented with a state-of-the-art cavity.

\section{Quantum state carving}

Counterfactual carving, recently proposed in \cite{Ramette2024_counterfactual}, is a heralded method that uses a high-finesse optical cavity to generate multipartite entangled states of atoms.  We can use counterfactual carving to generate high-fidelity resource states in a single step, without the time or physical qubit overheads typically required in this stage. 

To illustrate the concept, we consider $N$ three-level neutral atoms trapped in optical tweezers in the cavity mode, plus one identical source atom to herald the success or failure of the entanglement generation (Fig. \ref{fig:overview}b). Beginning in the coherent spin state $\ket{+}^{\otimes N}$ with all atomic spins aligned along the $x$ axis, we can ``carve" a GHZ entanglement in the $X$ basis by exponentially suppressing the unwanted odd Dicke levels of collective spin (Fig. \ref{fig:overview}a) \cite{Ramette2024_counterfactual}.  Applying a Hadamard gate to any one of the entangled qubits creates a locally-equivalent entangled state called a star graph \cite{Hein2006_stargraph}. A star graph is a type of resource state that can be visually represented as a ``central" qubit connected via C-phases to ``leaf" qubits (Fig. \ref{fig:overview}a), which provide the necessary overhead for constructing a large cluster state efficiently \cite{Lobl2024_losstolerant}.  Post-selected on heralded success, the infidelity of the resulting entanglement is 
\begin{equation}\label{infidelity}
\varepsilon_{\text{CF}} \sim (1/P_s)^{-C/N}  ,
\end{equation}
where $C$ is the single-atom cooperativity and $P_s$ is the probability of successfully carving the entanglement (see Appendix and Ref. \cite{Ramette2024_counterfactual}). 

In addition to its role as a `factory' for repeatedly generating star-type resource states via counterfactual carving, we can use the cavity to merge the resource states into the growing cluster state. We envision the central qubit of each resource state as the physical qubit to be added to the cluster state; we then attempt two-qubit C-phase gates between leaves of star graphs centered at neighboring lattice positions (Fig. \ref{fig:overview}c) \cite{Lobl2024_losstolerant}.  There are various schemes for implementing heralded non-local C-phase gates with a cavity, for example using counterfactual carving or other approaches \cite{Ramette2024_counterfactual,Sorensen2003_measurement,Kastoryano2011_dissipative,Welte2018_gate,Grinkemeyer2025}. Here we adopt a scheme introduced by Borregaard \textit{et al.} \cite{Borregaard2015_entanglement} that features a success probability scaling as $P_s \sim 1 - 6/\sqrt{C}$ at large $C$, as well as arbitrarily high post-selected fidelity.

If an attempt is heralded as a success, the two leaf qubits are measured in the $X$ basis to transfer the entanglement to the central qubits.  If it is heralded as a failure, the two leaf qubits are removed from the resource states via $Z$ measurements and the gate is attempted between two new leaf qubits.  Finally, if a star graph runs out of leaf qubits, the incomplete edges are marked as failures.  To deal with the qubits at the ends of a failed edge, we adopt an `adaptive' approach \cite{Auger2018_mbqc} in which one of the two qubits is measured in the $Z$ basis at random and contributes to the loss fraction. Assuming there are no errors on the physical qubits, the adaptive approach sets the threshold for entanglement failures at $14.5\%$ \cite{Auger2018_mbqc}. 

% rerun with 0.01 C.
% \onecolumngrid
% \begin{center}
\begin{figure*}[tb]
\makebox[\textwidth][c]{\includegraphics[width=1.1\textwidth]{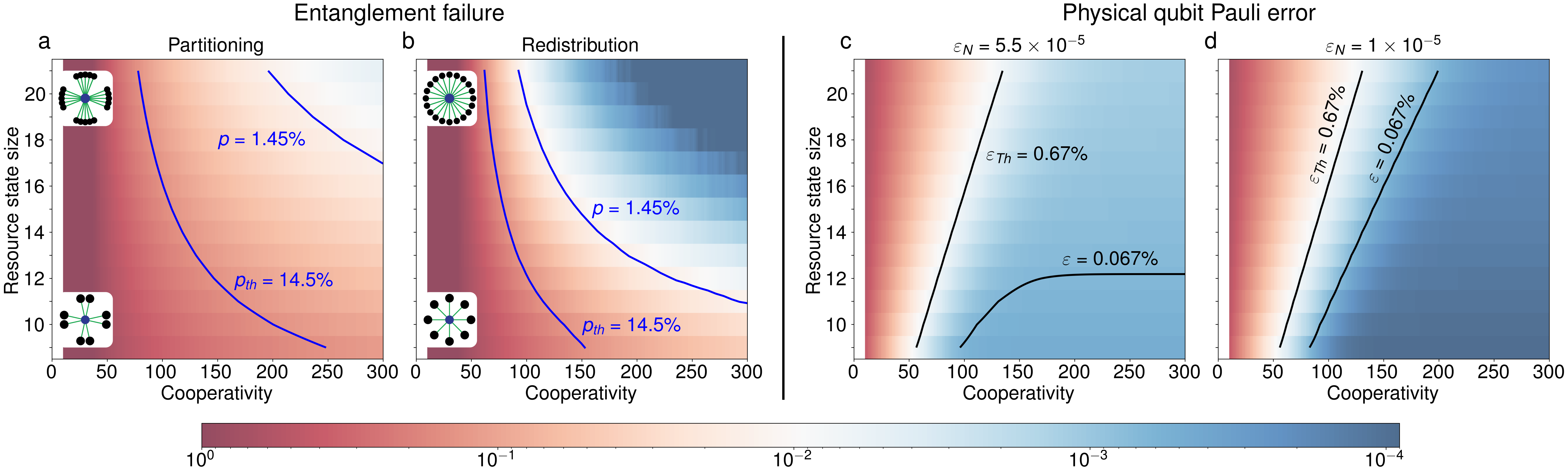}}
\caption{\textbf{Edge failure fraction and error rate in a distance-10 RHG cluster state. a, b:} Edge failure rates observed after the simulated construction of distance-10 RHG lattices, as a function of resource state size $N$ and cavity cooperativity $C$. For each $(C,N)$ combination, the edge failure rate in the partitioning method (\textbf{a.}) was calculated analytically using (\ref{eq:par}), while the edge failure rate in the redistribution method (\textbf{b.}) was simulated numerically.  Both methods assume an entangling gate failure probability of $6/\sqrt{C}$. The redistribution approach is necessary to achieve an edge failure fraction one order of magnitude below threshold ($p_{th}=14.5\%$) with cavity cooperativity and GHZ resource state size remaining within reasonable experimental limits.  \textbf{c, d:} The rate of physical qubit Pauli errors, which stem from the post-selected many-body infidelity of counterfactual carving as well as other non-carving error sources including measurement error and decoherence. When the non-carving errors reach a level of $5.5 \times 10^{-5}$ per qubit, reaching $0.067\%$ error rate is impossible for $N > 12$-qubit resource states. The intersection of $1.45\%$ edge failure under redistribution (second from left) and $0.067\%$ Pauli error rate outlines the parameter space corresponding to cluster states satisfying both thresholds by at least one order of magnitude (Fig. \ref{fig:intersections}).
\label{fig:levels}
}
\end{figure*}
% The downward slope of the level curves demonstrates that a larger cooperativity relaxes the necessary size of the resource states to achieve a given edge failure rate by improving the success probability of the cavity-mediated two-qubit gate. Still, 

\section{Methods}
We consider two methods for constructing edges between neighboring star graphs using a probabilistic gate.  The ``partitioning" method separates the leaf qubits into four groups corresponding to the four edges that connect a given physical qubit to its neighbors in the RHG cluster state. An edge in one of the four directions may only be attempted using leaf qubits from the designated group (Fig. \ref{fig:construction}c).  Once a gate is successful in a particular direction, any remaining leaf qubits in that group are removed by measurement.  If all the leaf qubits in the group are used up through a sequence of consecutive failures, the edge is deemed a failure.  Given a gate failure probability of $6/\sqrt{C}$ \cite{Borregaard2015_entanglement}, the expected edge failure rate for this method is
\begin{equation}\label{eq:par}
    p_{\text{edge}} = (6/\sqrt{C})^{\frac{N-1}{4}},
\end{equation}
where $N-1$ is the total number of available leaves in an $N$-qubit star graph and $C$ is the single-atom cooperativity.  From this we can directly compute the level curves of fixed $p_{\text{edge}}$ across the $(C,N)$ parameter space.

Alternatively, the more favorable ``redistribution" method allows for the leaves of a resource state to be allocated towards any of the four edges; unlike the partitioning method, there is no directional specification for any of the leaf qubits (Fig. \ref{fig:construction}b). A given edge is attempted until it is successful or either of the participating resource states runs out of available leaves. We use Monte Carlo simulation of this method to calculate the edge failure rate for various $(C,N)$ combinations.

We account for physical qubit Pauli errors introduced to the cluster state using a simple model that includes the post-selected infidelity of counterfactual carving, incorrect measurement outcomes during the heralded two-qubit gate, and qubit decoherence on a timescale of the $T_2$ dephasing time $\tau = 1.5s$ \cite{Bluvstein2022_transport}. 
%These sources of error all increase with the size of the resource states 
The total error caused by all these sources scales with the system size; however, we only consider large per-atom cooperativity ($C/N \gg 1$), meaning that measurement and decoherence errors ($\propto N$) dominate the overall error budget for large $N$. 
The error due to other sources does not scale with $N$. The total error is shown in Eq. \ref{errors}, with all errors which are proportional to $N$ grouped under the label $\varepsilon_N$.
 
\begin{equation}\label{errors}
     \varepsilon \approx \underbrace{e^{-\frac{8}{\pi^2}C/N}}_\text{CF carving} + N \underbrace{\bigg(\varepsilon_M + \frac{t}{\tau} + \dots\bigg)}_{\varepsilon_{N}} + \underbrace{( 4\;\varepsilon_G + ...)}_{\mathrm{other}}.
\end{equation}
Here $\varepsilon_M$ is the measurement infidelity.  A Pauli-Z error is introduced directly if there is an incorrect measurement result while removing the leaves from a resource state with four connections already made.  An effective Pauli error is introduced if the heralding operation incorrectly marks a failed two-qubit gate attempt as a success; the missing entanglement becomes a Pauli error during the computation. $\varepsilon_G$ is the post-selected infidelity of the heralded two-qubit gate stemming from experimental imperfections in the operation of the gate. The total error contribution will not exceed $4\varepsilon_G$ per physical qubit because each qubit makes at most four successful connections.

\begin{figure}[h!] 
\includegraphics[width=\linewidth]{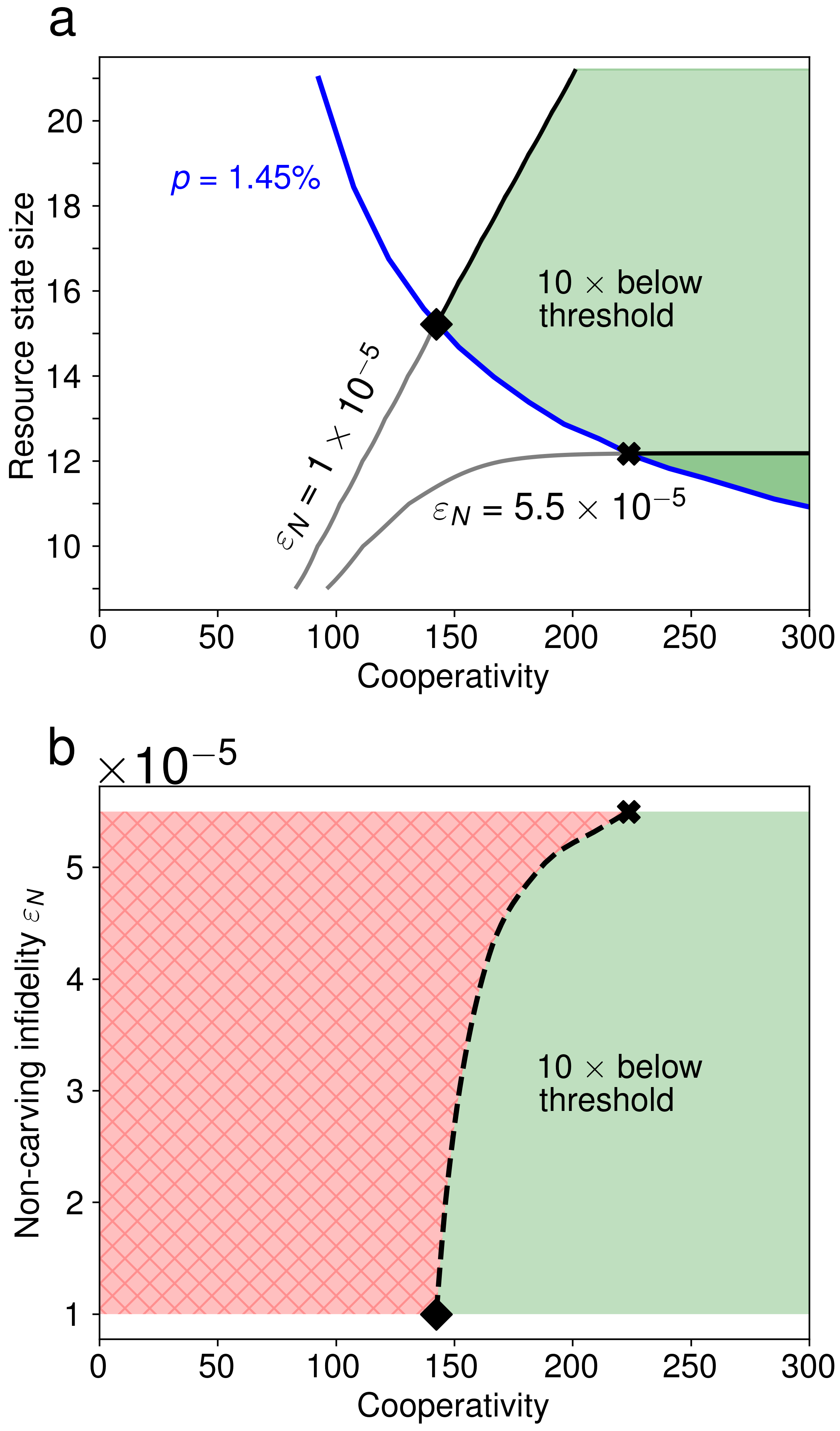}
\caption{\label{fig:intersections}\textbf{Parameter regimes for operating a factor of 10 below error correction thresholds. a.} Multiple level curves of $(C,N)$ coordinates that correspond to $1.45\%$ edge failure rate (blue) under the redistribution approach (Fig. \ref{fig:levels}b) and $0.067\%$ physical qubit Pauli error rate, both one order of magnitude below their respective thresholds. The gray lines correspond to two different non-carving infidelities $\varepsilon_N$, either $5.5\times 10^{-5}$ (black crossmark) or $1 \times 10^{-5}$ (black diamond). This infidelity contributes linearly with $N$ to the total Pauli error rate (\ref{errors}). \textbf{b.} Below $\varepsilon_N \approx 4 \times 10^{-5}$, further improvement in fidelity yields negligible relaxation in the required cooperativity to operate a factor of 10 below threshold.}
\label{fig:intersections}
\end{figure}

\section{Numerical results}
We simulated the construction of an RHG cluster state \cite{Tzitrin2021_flamingpy} with a code distance of 10, using both the redistribution and partitioning methods to compare performance in terms of entanglement failure rates in the resulting cluster state. This quantity can be directly calculated for the partitioning method using Eq. \eqref{eq:par}.  For the Monte Carlo simulation of the redistribution method, the likelihood of success of an entangling gate attempt between resource states is determined by the cooperativity. We performed calculations for resource state sizes ranging from 9 to 21 qubits, and for cooperativities between 10 and 300.

We begin by exploring the impact of construction methods on the requirements for achieving sub-threshold entanglement failure rates in the RHG lattice. The downward slope of the level curves in Figs. \ref{fig:levels}a,b demonstrates that in general, a larger cooperativity reduces the size of the resource states necessary to achieve a given edge failure rate by improving the success probability of the cavity-mediated two-qubit gate.  
%confirms the intuition that increasing the size of the resource states allows lower two-qubit heralded gate success probability, and by extension, cavity cooperativity.
Relative to the partitioning method, the redistribution approach, with its ability to dynamically assign leaves to different edges, yields a significant relaxation in both the cooperativity and resource state size necessary to achieve a particular suppression of the entanglement failure rate. For example, redistribution could enable a cavity with a cooperativity of 130 to achieve an entanglement failure rate beyond one order of magnitude below threshold with a resource state size of 17 qubits; in contrast, the partitioning method requires a much larger resource state size to achieve similar performance.
%The performance improvement arises because the redistribution method leverages all leaf qubits as overhead for any entanglement direction at a given node of the cluster state, whereas the partitioning method limits the overhead for each connection to only the leaf qubits in a specific group.

Fig. \ref{fig:levels}c,d illustrates the fundamental trade-off between the entanglement failure rate and the physical qubit error rate in constructing the RHG lattice.  Although carving larger resource states lowers the chance of a failed edge, for fixed cooperativity the fidelity of the output resource state decreases with size (\ref{infidelity}).  This introduces Pauli errors to the RHG lattice when the star graphs are merged together.  Importantly, the ``non-carving" infidelity $\varepsilon_n$, representing the combination of all other error sources that scale to first order with $N$ (decoherence, measurement errors, etc.) can also overwhelm the counterfactual carving infidelity.  In Fig. \ref{fig:levels}c we show that, above a certain resource state size, a non-carving infidelity of $5.5 \times 10^{-5}$ becomes the dominant contributor to the overall physical qubit error rate; further improvements in cooperativity do not reduce the error rate beyond one order of magnitude below threshold.  That non-carving error sources at a level of $10^{-5}$ become limiting is a testament to the enabling role played by counterfactual carving in these results. Prior carving proposals with infidelities that scale as $(C/N)^{-1}$ 
\cite{Chen2015_carving} require inconveniently large cooperativities to achieve comparable physical qubit error rates.

Fig. \ref{fig:intersections} explores the intersection between level curves of constant edge failure and physical qubit Pauli errors to identify a target cooperativity and resource state size that enable sub-threshold performance while remaining achievable in state-of-the-art experiments. Fig. \ref{fig:intersections}a demonstrates how error-prone measurement operations and high decoherence would require excessively large cooperativity. However, we find that limiting non-carving infidelity $\varepsilon_N$ to below $4 \times 10^{-5}$ is also unnecessary, as the corresponding cooperativity for operating $10 \;\times$ below threshold decreases only slightly (Fig. \ref{fig:intersections}b).  This is realistic for near-term hardware given recent experimental demonstrations of cavity-assisted measurement infidelity at a level of $10^{-3}$ \cite{Reichel2010_readout} and coherent, parallel transport of neutral atoms across large distances on timescales $< 10^{-3} \times T_2$ \cite{Bluvstein2022_transport}.  Moreover, cavity cooperativities greater than 100 have been experimentally realized \cite{Colombe2007_cooperativity}. Our results suggest that at a non-carving infidelity of $\varepsilon_N=3\times 10^{-5}$ per qubit, a cavity with cooperativity of 160 generating GHZ states of 15 atoms (Fig. \ref{fig:intersections}a), could be used to construct a cluster state with losses and Pauli errors 10 $\times$ below their respective thresholds, enabling unprecedented logical success rates.

\section{Conclusion}

Existing proposals for neutral-atom quantum computing platforms typically employ deterministic Rydberg-based entangling gates to implement error correction through circuit-based or measurement-based models \cite{Evered2023_gate,Sahay2023_erasure}. Here we propose an integrated atom-cavity architecture as an alternative pathway to fault-tolerance through probabilistic, heralded gates and measurement-based quantum processing. This approach retains the attractive properties of neutral-atom qubits, including high-fidelity single-qubit gates, long coherence times, dynamic reconfigurability, and excellent readout fidelity, while leveraging the strong atom-cavity interaction to generate and connect large resource states with high fidelity. Notably, our approach can be implemented with near-term experimental capabilities \cite{Colombe2007_cooperativity,Reichel2010_readout}, enabling the generation of a fault-tolerant cluster state with losses and Pauli errors well below the respective thresholds.

\section{Acknowledgements}
We thank Shayan Majidy for comments on the manuscript.  L.M.S. acknowledges support from the MIT UROP program. 
G.B. acknowledges support from the MIT Patrons of Physics Fellows Society.
This project was funded in part by DARPA under the ONISQ program (grant \# W911NF2010021), the US Department of Energy (Quantum Systems Accelerator, grant \# DE-AC02-05CH11231), the Center for Ultracold Atoms (an NSF Physics Frontiers Center, Grant \# PHY- 1734011), QuSeC-TAQS from NSF (grant \# 2326787), (grant \# W911NF2320219), the Army Research Office (grant \# W911NF2010082), IARPA under the ELQ program (grant \# W911NF2320219), and QuEra Computing.

\subsection{Author Contributions}

J.R. and J.S. conceived of the original idea. G.B. and L.M.S. developed the methods.  L.M.S. performed the numerical analysis and drafted the manuscript.  All authors discussed the method and results and contributed to the manuscript.  V.V. supervised the project.

\section{Appendix}
\subsection{Counterfactual Carving}

\begin{figure}[h!]
\includegraphics[width=\linewidth]{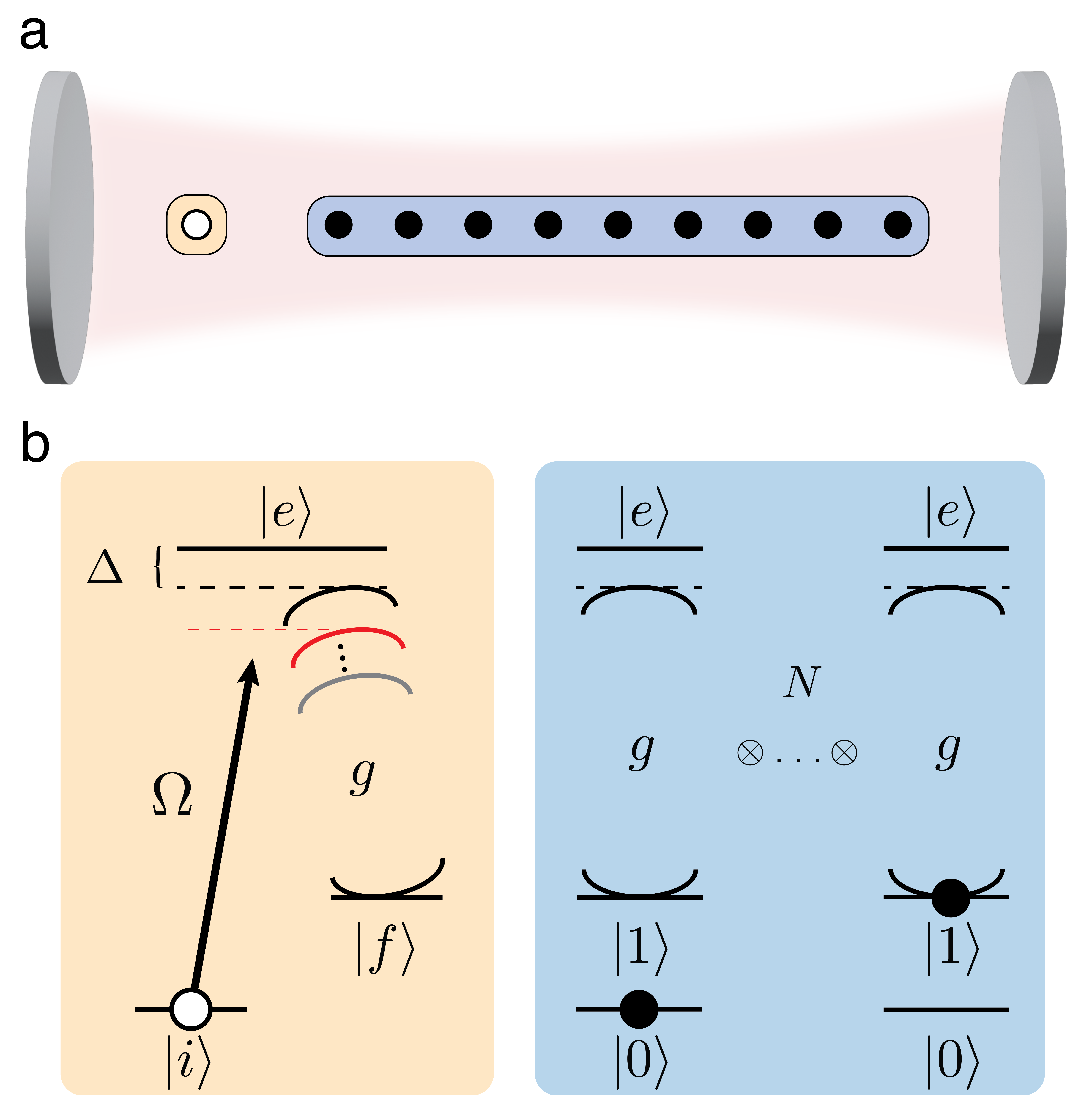}
\caption{Cavity layout (\textbf{a.}) and atomic level scheme (\textbf{b.}) for counterfactual carving \cite{Ramette2024_counterfactual}. A source atom (beige) is used to probe the cavity resonance in the dispersive limit of a strongly-coupled ensemble of atoms (blue).  By heralding on attempts where the source atom does not undergo a transition from $\ket{i}$ to $\ket{f}$, multipartite entanglement in the ensemble can be produced in a heralded manner.
}
\label{fig:cfcarving}
\end{figure}

We summarize here some of the details of counterfactual carving and the contours of an implementation using atomic qubits trapped in optical tweezers in a high-finesse optical cavity.  An in-depth explanation and analysis of counterfactual carving can be found in Ref. \cite{Ramette2024_counterfactual}.

We consider each atom to have two long-lived qubit states; one state ($\ket{1}$) is coupled via the cavity to an excited state with atom-cavity coupling $g$ and detuning $\Delta$ (Fig. \ref{fig:cfcarving}).
In the large detuning regime, where $\Delta \gg g$, for each of the $N$ ensemble atoms in the cavity-coupled qubit state $\ket{1}$, the cavity resonance is shifted by $g^2/\Delta$ (dispersive limit). An additional three-level ``source" atom coupled to the cavity mode acts as a single-photon source for probing the cavity resonance and allows for high-fidelity heralding of the entangling operation.  The figure of merit capturing the strength of the single-atom coupling to the cavity mode is the cooperativity $C \equiv g^2/\kappa \Gamma$, where $2g$ is the vacuum Rabi frequency, $\kappa$ is the cavity photon decay rate and $\Gamma$ is the atomic excited-state population decay rate.

The shifts of the cavity resonance in the large-detuning regime discriminate between eigenstates $\ket{m},\; m \in [-N/2,N/2]$ of the collective spin of the atomic ensemble, known as Dicke levels.  Beginning from the CSS $\ket{+}^{\otimes N} = \sum c_m \ket{m}$, the GHZ state in the $x$ basis consists of a superposition of the even-$m$ Dicke levels.

By addressing the source atom with different frequencies (with coupling $\Omega$) tuned to various resonance shifts of the cavity and then observing the state of the source atom after some time $t$, we project the original ensemble wavefunction into a GHZ state in the $x$ basis by exponentially suppressing the odd-$m$ weights relative to the even-$m$ levels.  The duration of the source tones determines the probability of success of the carving operation, while the infidelity of the entanglement, after post-selection on the source atom remaining in state $\ket{i}$, scales as 
\begin{equation}
    \varepsilon_{\text{CF}} \sim e^{-\frac{8}{\pi^2}\frac{C}{N}} \;
\end{equation}
for $O(1)$ success probability.  The post-selected infidelity can be made arbitrarily small in exchange for longer wait times and lower success probability.

\bibliography{apsform_fullauthors}% Produces the bibliography via BibTeX.

\end{document}